\documentclass[11pt]{cernrep}
\usepackage{graphicx}
\usepackage{cite,./mcite}
\newcommand{\as}{\alpha_s}
\begin{document}
\title{A posteriori inclusion of PDFs in NLO QCD final-state calculations\footnote{\hspace{2ex}
Contribution to the CERN - DESY Workshop 2004/2005, {\it HERA and
the LHC}.}
}
\author{Tancredi Carli$^{1}$, Gavin P. Salam$^{2}$, Frank Siegert$^{1}$}
\institute{
$^{1}$ CERN, Department of Physics, CH-1211 Geneva 23, Switzerland \\
$^{2}$ LPTHE, Universities of Paris VI and VII and CNRS, 75005, Paris, France
}
\maketitle

\begin{abstract}
  Any NLO calculation of a QCD final-state observable involves Monte
  Carlo integration over a large number of events. For DIS and hadron
  colliders this must usually be repeated for each new PDF set, making
  it impractical to consider many `error' PDF sets, or carry out PDF
  fits.
  Here we discuss ``a posteriori'' inclusion of PDFs, whereby the
  Monte Carlo run calculates a grid (in $x$ and $Q$) of cross section
  weights that can subsequently be combined with an arbitrary PDF.
  The procedure is numerically equivalent to using an interpolated
  form of the PDF. The main novelty relative to prior work is the use
  of higher-order interpolation, which substantially improves the
  tradeoff between accuracy and memory use.
  An accuracy of about $0.01$\% has been reached for the single
  inclusive cross-section in the central rapidity region $|y|<0.5$ for
  jet transverse momenta from $100$ to $5000 \mathrm{GeV}$.
  This method should facilitate the consistent inclusion of
  final-state data from HERA, Tevatron and LHC data in PDF fits, thus
  helping to increase the sensitivity of LHC to deviations from
  standard Model predictions.

\end{abstract}

\section{Introduction}
The Large Hadron Collider (LHC), currently under construction at CERN,
will collide protons on protons with an energy of $7$~{\rm TeV}.
Together with its high collision rate the high available centre-of-mass
energy 
will make it possible to test new interactions at very short
distances that might be revealed
in the production cross-sections of Standard Model (SM) particles at very
high transverse momentum ($P_T$)
as deviation from the SM theory.

The sensitivity to new physics crucially depends on experimental uncertainties in the measurements
and on theoretical uncertainties in the SM predictions. It is therefore important to work out
a strategy to minimize both the experimental and theoretical uncertainties from LHC data.
For instance, one could use single inclusive jet  or Drell-Yan
cross-sections at low $P_T$ to constrain the PDF uncertainties at high $P_T$.
Typical residual renormalisation and factorisation scale uncertainties
in next-to-leading order (NLO) calculations for single inclusive
jet-cross-section are about $5-10\%$ and should hopefully be reduced
as NNLO calculations become available. The impact of PDF uncertainties on the other
hand can be substantially larger in some regions, especially at large $P_T$,
and for example at $P_T = 2000$~{\rm GeV} dominate the overall
uncertainty of $20\%$.
%
If a suitable combination of data measured
at the Tevatron and LHC can be included in global NLO QCD analyses,
the PDF uncertainties can be constrained.

The aim of this contribution is to propose a method for consistently
including final-state observables in global QCD analyses.

For inclusive data like the proton structure function $F_2$ in deep-inelastic
scattering (DIS) the perturbative coefficients
are known analytically. During the fit the cross-section
can therefore be quickly calculated from the strong coupling ($\as$) and the PDFs
and can be compared to the measurements.
However, final state observables, where detector acceptances
or jet algorithms are involved in the definition of the perturbative coefficients
(called ``weights'' in the following),
have to be calculated using NLO Monte Carlo programs. Typically such programs
need about one day of CPU time to calculate accurately the cross-section.
It is therefore necessary to find a way to calculate the perturbative
coefficients with high precision in a long run and to include
 $\as$ and the PDFs  ``a posteriori''.

To solve this problem
many methods have been proposed in the past
\cite{Graudenz:1995sk,Kosower:1997vj,Stratmann:2001pb,wobisch,wobisch,zeus2005}.  In principle the
highest efficiencies can be obtained by taking moments with respect to
Bjorken-$x$ \cite{Graudenz:1995sk,Kosower:1997vj}, because this
converts convolutions into multiplications. This can have notable
advantages with respect to memory consumption, especially in cases
with two incoming hadrons.  On the other hand, there are complications
such as the need for PDFs in moment space and the
associated inverse Mellin transforms.

Methods in $x$-space have traditionally been somewhat less efficient,
both in terms of speed (in the `a posteriori' steps --- not a major
issue here) and in terms
of memory consumption. They are, however, somewhat more transparent
since they provide direct information on the $x$ values of relevance.
Furthermore they can be used with any PDF.
The use of $x$-space methods can be further improved by using
methods developed originally for PDF evolution \cite{Ratcliffe:2000kp,Dasgupta:2001eq}.
\section{PDF-independent representation of cross-sections}
\subsection{Representing the PDF on a grid}
We make the assumption that PDFs can be accurately represented by storing their values on a
two-dimensional grid of points and using $n^{\mathrm{th}}$-order interpolations between those points.
Instead of using the parton momentum fraction $x$ and the factorisation scale $Q^2$, we use a
variable transformation that provides good coverage of the full $x$
and $Q^2$ range
with uniformly spaced grid points:%
\footnote{An alternative for the $x$ grid is to use $y = \ln 1/x +
  a(1-x)$ with $a$ a parameter that serves to increase the density of
  points in the large $x$ region.}
\begin{equation}
\label{eq:ytau}
y(x) = \ln \frac{1}{x} \; \; \; {\rm and} \; \; \;
\tau(Q^2) = \ln \ln \frac{Q^2}{\Lambda^2}.
\end{equation}
The parameter $\Lambda$ is to be chosen of the order of $\Lambda_{\mathrm{QCD}}$, but not necessarily identical.
The PDF $q(x,Q^2)$ is then represented by its values $q_{i_y,i_\tau}$ at the 2-dimensional
grid point $(i_y \, \delta y, i_\tau \, \delta \tau)$, where $\delta
y$ and $\delta \tau$ denote the grid spacings,
and obtained elsewhere by interpolation:
\begin{equation}
\label{eq:interp}
q(x,Q^2) = \sum_{i=0}^n \sum_{\iota=0}^{n'} q_{k+i,\kappa+\iota} \,\,
I_i^{(n)}\left(  \frac{y(x)}{\delta y}  - k \right)\,
I_\iota^{(n')}\left(  \frac{\tau(Q^2)}{\delta\tau}-\kappa  \right),
\end{equation}
where $n$, $n'$ are the interpolation orders.
The interpolation function $I_i^{(n)}(u)$ is 1 for $u=i$ and otherwise is given by:
\begin{equation}
\label{eq:Ii}
I_i^{(n)}(u) = \frac{(-1)^{n-i}}{i!(n-i)!} \frac{u (u-1) \ldots (u-n)}{u-i}\,.
\end{equation}
Defining $\mathrm{int} (u)$ to be the largest integer such that $\mathrm{int}(u) \le u$,
$k$ and $\kappa$ are defined as:
\begin{eqnarray}
\label{eq:kchoice}
k(x) =& \mathrm{int} \left( \frac{y(x)}{\delta y} - \frac{n-1}{2} \right), &
\kappa(x) = \mathrm{int} \left( \frac{\tau(Q^2)}{\delta \tau} - \frac{n'-1}{2} \right).
\end{eqnarray}
%
Given finite grids whose vertex indices range from $0\ldots N_y-1$ for
the $y$ grid and $0\ldots N_\tau-1$ for the $\tau$ grid, one should
additionally require that eq.~(\ref{eq:interp}) only uses available
grid points. This can be achieved by remapping $k \to
\max(0,\min(N_y-1-n,k))$ and $\kappa \to
\max(0,\min(N_\tau-1-n',\kappa))$.

\subsection{Representing the final state cross-section weights on a
  grid (DIS case)}
Suppose that we have an NLO Monte Carlo program that produces events $m=1\dots N$.
Each event $m$ has an $x$ value, $x_m$, a $Q^2$ value, $Q^2_m$, as well as a weight, $w_m$,
and a corresponding order in $\as$, $p_m$.
Normally one would obtain the final result $W$ of the Monte Carlo integration from:\footnote{Here, and in the following,
renormalisation and factorisation scales have been set equal for simplicity.}
\begin{equation}
\label{eq:normalint}
 W = \sum_{m=1}^N \,w_m \, \left( \frac{\alpha_s(Q_m^2)} {2\pi}\right)^{p_m}  \, q(x_m,Q^2_m).
\end{equation}

Instead one introduces a weight grid $W_{i_y,i_\tau}^{(p)}$ and then for each event updates
a portion of the grid with:\\
$i = 0\dots n,\; \iota = 0\dots n':$
\begin{eqnarray}
\label{eq:weight2evolve}
W_{k+i,\kappa + \iota}^{(p_m)} \to W_{k+i,\kappa + \iota}^{(p_m)} + w_m\,
  I_i^{(n)} \left(\frac{y(x_m)}{\delta y} - k\right)
  I_{\iota}^{(n')}\left(\frac{\tau(Q^2_m)}{\delta \tau} - \kappa \right), \\
 \;\;\; {\rm where} \;\;\;
  k \equiv k(x_m),\; \kappa \equiv \kappa(Q^2_m). \nonumber
  \end{eqnarray}
The final result for $W$, for an arbitrary PDF, can then be obtained \emph{subsequent}
to the Monte Carlo run:
\begin{equation}
\label{eq:WfinalxQ}
W = \sum_p \sum_{i_y} \sum_{i_\tau}
W_{i_y,i_\tau}^{(p)} \, \left( \frac{\alpha_s\left({Q^2}^{(i_\tau)}\right)}{2\pi}\right)^{p} q \!\left(x^{(i_y)}, {Q^2}^{(i_\tau)} \right)\,,
\end{equation}
where the sums index with $i_y$ and $i_\tau$ run over the number of grid points and
we have have explicitly introduced $x^{(i_y)}$ and ${Q^2}^{(i_\tau)}$ such that:
\begin{equation}
\label{eq:xQdefs}
 y(x^{(i_y)}) = i_y \, \delta y \quad {\rm and} \quad
\tau\left({Q^2}^{(i_\tau)}\right) =  i_\tau \, \delta \tau.
\end{equation}

\subsection{Including renormalisation and factorisation scale dependence }
If one has the weight matrix $W_{i_y,i_\tau}^{(p)}$ determined separately order by order in
$\as$, it is straightforward to vary the renormalisation $\mu_R$  and
factorisation $\mu_F$ scales a posteriori (we assume that they were kept equal
in the original calculation).

It is helpful to introduce some notation relating to the DGLAP evolution 
equation:
\begin{equation}
  \label{eq:DGLAP}
  \frac{d q(x,Q^2)}{d \ln Q^2} = \frac{\alpha_s(Q^2)}{2\pi} (P_0 \otimes q)(x,Q^2)
                + \left(\frac{\alpha_s(Q^2)}{2\pi}\right)^2 (P_1
                \otimes q)(x,Q^2) + \ldots,
\end{equation}
where the $P_0$ and $P_1$ are the LO and NLO matrices of DGLAP
splitting functions that operate on vectors (in flavour space) $q$ of
PDFs.
Let us now restrict our attention to the NLO case where we have just
two values of $p$, $p_{\mathrm{LO}}$ and $p_{\mathrm{NLO}}$.
Introducing $\xi_R$ and $\xi_F$ corresponding to the factors by which one varies $\mu_R$ and $\mu_F$ respectively,
for arbitrary $\xi_R$ and $\xi_F$ we may then write:
\begin{eqnarray}
  \label{eq:Wfinalxi}
  W(\xi_R, \xi_F) = \sum_{i_y} \sum_{i_\tau}
    \left(\frac{\alpha_s\left(\xi_R^2 {Q^2}^{(i_\tau)}\right)\,
               }{2\pi}
     \right)^{p_{\mathrm{LO}}}
  W_{i_y,i_\tau}^{(p_{\mathrm{LO}})}
   q \!\left(x^{(i_y)}, \xi_F^2  {Q^2}^{(i_\tau)} \right) +
   \nonumber \\
  \left(\frac{\alpha_s\left(\xi_R^2 {Q^2}^{(i_\tau)} \right)\,
             }{2\pi}
   \right)^{p_{\mathrm{NLO}}}
  \left[
    \left( W_{i_y,i_\tau}^{(p_{\mathrm{NLO}})} + 2\pi  \beta_0 p_{\mathrm{LO}} \ln \xi_R^2
         \,W_{i_y,i_\tau}^{(p_{\mathrm{LO}})}
    \right)  q \!\left(x^{(i_y)}, \xi_F^2 {Q^2}^{(i_\tau)} \right)
    \right. \\\left.
     - \ln \xi_F^2 \,W_{i_y,i_\tau}^{(p_{\mathrm{LO}})}
     (P_0\otimes q) \!\left(x^{(i_y)}, \xi_F^2 {Q^2}^{(i_\tau)} \right)
  \right] \,, \nonumber
\end{eqnarray}
where $\beta_0 = (11 N_c - 2n_f)/(12\pi)$ and $N_c$ ($n_f$) is the number of colours (flavours).
Though this formula is
given for $x$-space based approach, a similar formula applies for
moment-space approaches. Furthermore it is straightforward to extend
it to higher perturbative orders.

 \subsection{Representing the weights in the case of two incoming hadrons}
In hadron-hadron scattering one can use analogous procedures with one more dimension.
Besides $Q^2$, the weight grid depends on the momentum fraction of the first ($x_1$) and
second ($x_2$) hadron.

In the case of jet production in proton-proton collisions
the weights generated by the Monte Carlo program as well as the PDFs
can be organised in seven possible initial state combinations of partons:
\begin{eqnarray}
\mathrm{gg}: \;\; F^{(0)}(x_{1}, x_{2}; Q^{2}) &=& G_{1}(x_{1})G_{2}(x_{2})\\
\mathrm{qg}: \;\; F^{(1)}(x_{1}, x_{2}; Q^{2}) &=& \left(Q_{1}(x_{1})+
                  \overline Q_{1}(x_{1})\right) G_{2}(x_{2})\\
\mathrm{gq}: \;\; F^{(2)}(x_{1}, x_{2}; Q^{2}) &=&  G_{1}(x_{1})\left(Q_{2}(x_{2})+
                  \overline Q_{2}(x_{2})\right)\\
\mathrm{qr}: \;\; F^{(3)}(x_{1}, x_{2}; Q^{2}) &=&  Q_{1}(x_{1}) Q_{2}(x_{2})
                                        + \overline Q_{1}(x_{1}) \overline Q_{2}(x_{2}) -D(x_{1}, x_{2})\\
\mathrm{qq}: \;\; F^{(4)}(x_{1}, x_{2}; Q^{2}) &=& D(x_{1}, x_{2})\\
\mathrm{q\bar q}: \;\; F^{(5)}(x_{1}, x_{2}; Q^{2}) &=& \overline D(x_{1}, x_{2})\\
\mathrm{q\bar r}: \;\; F^{(6)}(x_{1}, x_{2}; Q^{2}) &=& Q_{1}(x_{1}) \overline Q_{2}(x_{2})
                   + \overline Q_{1}(x_{1}) Q_{2}(x_{2})       -\overline D(x_{1}, x_{2}),
\end{eqnarray}
where $g$ denotes gluons, $q$ quarks and $r$ quarks of different flavour $q \neq r$
and we have used the generalized PDFs defined as:
\begin{eqnarray}
 G_{H}(x) = f_{0/H}(x,Q^{2}), &&
 Q_{H}(x) = \sum_{i = 1}^{6} f_{i/H}(x,Q^{2}), \;\;
 \overline Q_{H}(x) = \sum_{i = -6}^{-1} f_{i/H}(x,Q^{2}), \nonumber \\
 D(x_{1}, x_{2}) &=& \mathop{\sum_{i = -6}^{6}}_{i\neq0} f_{i/H_1}(x_{1},Q^2) f_{i/H_2}(x_{2},Q^{2}), \\
 \overline D(x_{1}, x_{2}, \mu^{2}_{F}) &=&
  \mathop{\sum_{i = -6}^{6}}_{i\neq0} f_{i/H_1}(x_{1},Q^{2}) f_{-i/H_2}(x_{2},Q^{2}), \nonumber \;\;
\end{eqnarray}
where $f_{i/H}$ is the PDF of flavour $i=-6 \dots 6$ for hadron $H$
and $H_1$ ($H_2$) denotes the first or second hadron\footnote{
In the above equation we follow the standard PDG Monte Carlo numbering
scheme \cite{Eidelman:2004wy} where gluons
are denoted as $0$, quarks have values from $1$-$6$ and anti-quarks have the corresponding
negative values.}.

The analogue of eq.~\ref{eq:WfinalxQ} is then given by:
\begin{equation}
\label{eq:WfinalxQ_twohadrons}
W = \sum_p \sum_{l=0}^{6} \sum_{i_{y_1}} \sum_{i_{y_2}} \sum_{i_\tau}
W_{i_{y_1},i_{y_2},i_\tau}^{(p)(l)} \, \left( \frac{\alpha_s\left({Q^2}^{(i_\tau)}\right)}{2\pi}\right)^{p}
F^{(l)}\left(x_1^{(i_{y_1})}, x_2^{(i_{y_1})},  {Q^2}^{(i_\tau)}\right).
\end{equation}

 \subsection{Including scale depedence in the case of two incoming hadrons}
It is again possible to choose arbitrary renormalisation and
factorisation scales, specifically for NLO accuracy:
\begin{eqnarray}
  \label{eq:Wfinalxi_twohadrons}
  W(\xi_R, \xi_F) = \sum_{l=0}^{6} \sum_{i_{y_1}} \sum_{i_{y_2}} \sum_{i_\tau}
    \left(\frac{\alpha_s\left(\xi_R^2 {Q^2}^{(i_\tau)}\right)\,
               }{2\pi}
     \right)^{p_{\mathrm{LO}}}
     W_{i_{y_1},i_{y_2},i_\tau}^{(p_{\mathrm{LO}})(l)}
  F^{(l)}\left(x_1^{(i_{y_1})}, x_2^{(i_{y_1})}, \xi_F^2{Q^2}^{(i_\tau)}\right)
    +
   \nonumber \\
  \left(\frac{\alpha_s\left(\xi_R^2 {Q^2}^{(i_\tau)} \right)\,
             }{2\pi}
   \right)^{p_{\mathrm{NLO}}}
  \left[
    \left(
      W_{i_{y_1},i_{y_2},i_\tau}^{(p_{\mathrm{NLO}})(l)}
      + 2\pi  \beta_0 p_{\mathrm{LO}} \ln \xi_R^2
         \,
         W_{i_{y_1},i_{y_2},i_\tau}^{(p_{\mathrm{LO}})(l)}
    \right)
    F^{(l)}\left(x_1^{(i_{y_1})}, x_2^{(i_{y_1})}, \xi_F^2{Q^2}^{(i_\tau)}\right)
    \right. \\\left.
     - \ln \xi_F^2 \,
     W_{i_{y_1},i_{y_2},i_\tau}^{(p_{\mathrm{LO}})(l)}
     \left(
     F^{(l)}_{q_1 \to P_0\otimes q_1}\left(x_1^{(i_{y_1})}, x_2^{(i_{y_1})},
       \xi_F^2{Q^2}^{(i_\tau)}\right) +
     F^{(l)}_{q_2 \to P_0\otimes q_2}\left(x_1^{(i_{y_1})}, x_2^{(i_{y_1})},
       \xi_F^2{Q^2}^{(i_\tau)}\right)
     \right)
  \right] \,, \nonumber
\end{eqnarray}
where $F^{(l)}_{q_1 \to P_0\otimes q_1}$ is calculated as $F^{(l)}$,
but with $q_1$ replaced wtih $P_0 \otimes q_1$, and analogously for
$F^{(l)}_{q_2 \to P_0\otimes q_2}$.

\section{Technical implementation}

To test the scheme discussed above we use the NLO Monte Carlo program
NLOJET++ \cite{Nagy:2003tz,*Nagy:2001fj,*Nagy:2001xb} and the CTEQ6 PDFs \cite{Pumplin:2002vw}.
The grid $W_{i_{y_1},i_{y_2},i_\tau}^{(p)(l)}$ of eq.~\ref{eq:WfinalxQ_twohadrons}
is filled in a NLOJET++ user module. This module has access to the event
weight and parton momenta and it is here that one specifies and calculates the physical
observables that are being studied (e.g. jet algorithm).

Having filled the grid we construct the cross-section
in a small standalone program which reads the weights from the grid
and multiplies them with an arbitrary  $\as$ and
PDF  according to eq.~\ref{eq:WfinalxQ_twohadrons}.
This program  runs very fast (in the order of seconds) and can be called
in a PDF fit.

The connection between these two programs is accomplished via a C++ class, which provides methods e.g.
for creating and optimising the grid, filling weight events and saving it to disk.
The classes are general enough to be extendable for the use with other NLO calculations.

The complete code for the NLOJET++ module, the C++ class and the standalone job is
available from the authors.
It is still in a development, testing and tuning stage, but help and more ideas are welcome.

\subsection{The C++ class}
\label{sec:class}
The main data members of this class are the grids implemented as
arrays of three-dimensional ROOT histograms, with each grid point at the bin centers\footnote{
ROOT histograms are easy to implement, to represent and to manipulate.
They are therefore ideal in an early development phase.
An additional
advantage is the automatic file compression to save space.
The overhead of storing some empty bins is largely reduced by optimizing the
$x_1$, $x_2$ and $Q^2$ grid boundaries using the NLOJET++ program before final filling.
To avoid this residual overhead and to exploit certain symmetries in the grid, a special data
class (e.g. a sparse matrix) might be constructed in the future.}:
\begin{equation}
 {\rm TH3D[p][l][iobs](x_1,x_2,Q^2)},
\end{equation}
where the $l$ and $p$ are explained  in eq.~\ref{eq:WfinalxQ_twohadrons}
and $iobs$ denotes the observable bin, e.g. a given $P_T$ range\footnote{
For the moment we construct a grid for each initial state parton configuration. It will be easy
to merge the $qg$ and the $gq$ initial state parton configurations in one grid.
In addition, the weights for some of the initial state parton configurations
are symmetric in $x_1$ and $x_2$.
This could be exploited in future applications to further reduce the grid size.
}.

The  C++ class initialises, stores and fills the grid using the following main
methods:
\begin{itemize}
\item \emph{Default constructor:} Given the pre-defined kinematic regions of interest, it initializes the grid.
\item \emph{Optimizing method:} Since in some bins the weights will be zero over a large
                                     kinematic region in $x_1, x_2, Q^2$,
                                     the optimising method implements an automated procedure to adapt the
                                     grid boundaries for each observable bin.
                                     These boundaries are calculated in a first (short) run.
                                     In the present implementation, the optimised grid has a fixed
                                     number of grid points. Other choices, like
                                     a fixed grid spacing, might be implemented in the future.
\item \emph{Loading method:} Reads the saved weight grid from a ROOT file
\item \emph{Saving method:} Saves the complete grid to a ROOT file, which will be automatically compressed.
\end{itemize}

 \subsection{The user module for NLOJET++}
The user module has to be adapted specifically to the exact definition of the cross-section calculation.
If a grid file already exists in the directory where NLOJET++ is started,
the grid is not started with the default constructor, but with the optimizing method (see \ref{sec:class}).
In this way the grid boundaries are optimised for each observable bin.
This is necessary to get very fine grid spacings without exceeding the computer memory.
The grid is filled at the same place where the standard NLOJET++ histograms are filled.
After a certain number of events, the grid is saved in a root-file and the calculation
is continued.

\subsection{The standalone program for constructing the cross-section}
The standalone program calculates the cross-section in the following way:
\begin{enumerate}
\item Load the weight grid from the ROOT file
\item Initialize the PDF interface\footnote{We use the C++ wrapper
of the LHAPDF interface \cite{Whalley:2005nh}.}, load $q(x,Q^2)$ on a
helper PDF-grid (to increase the performance)
\item For each observable bin, loop over $i_{y_1},i_{y_2}, i_\tau, l, p$
      and calculate $F^{l}(x_1, x_2,Q^2)$ from the appropriate PDFs $q(x,Q^2)$,
      multiply  $\as$
      and the weights from the grid
      and sum over the initial state parton configuration $l$,
      according to eq.~\ref{eq:WfinalxQ_twohadrons}.
\end{enumerate}

\section{Results}
\label{sec:results}
We calculate the single inclusive jet cross-section as a function of the jet transverse
momentum ($P_T$) for jets within a rapidity of $|y|<0.5$. To define the jets we use
the seedless cone jet algorithm as implemented in NLOJET++ using the four-vector recombination
scheme and the midpoint algorithm. The cone radius has been put to $R=0.7$, the overlap fraction
was set to $f=0.5$. We set the renormalisation and factorization scale to $Q^2=P_{T,max}^2$,
where $P_{T,max}$ is the $P_T$ of the highest $P_T$ jet in the required rapidity region\footnote{
Note that beyond LO the $P_{T,max}$ will in general differ from the $P_T$
of the other jets, so when binning an inclusive jet cross section, the
$P_T$ of a given jet may not correspond to the renormalisation scale
chosen for the event as a whole. For this reason we shall need separate
grid dimensions for the jet $P_T$ and for the renormalisation scale. Only
in certain moment-space approaches \cite{Kosower:1997vj} has this
requirement so far been efficiently circumvented.}.

In our test runs, to be independent from statistical fluctuations
(which can be large in particular in the NLO case),
we fill in addition to the grid a reference histogram in the standard way
according to eq.~\ref{eq:normalint}.

The choice of the grid architecture depends on the required accuracy, on the exact cross-section definition
and on the available computer resources. Here, we will just sketch the influence of the grid architecture
and the interpolation method on the final result. We will investigate an example where we
calculate the inclusive jet cross-section in $N_{\mathrm{obs}} = 100$ bins in the kinematic range
$100\, \leq P_T \leq 5000\,\mathrm{GeV}$.
In future applications this can serve as guideline for a user to adapt the grid method to his/her specific problem.
We believe that the code is transparent and flexible enough to adapt to many applications.

As reference for comparisons of different grid architectures and interpolation methods
we use the following:
\begin{itemize}
\item \emph{Grid spacing in $y(x)$:} $10^{-5} \leq x_1, x_2 \leq 1.0$ with $N_y=30$
\item \emph{Grid spacing in $\tau(Q^2)$:} $100\,\mathrm{GeV} \leq Q \leq 5000\,\mathrm{GeV}$ with $N_\tau=30$
\item \emph{Order of interpolation:} $n_y=3,\, n_\tau=3$
\end{itemize}
The grid boundaries correspond to the user setting for the first run which determines
the grid boundaries for each observable bin.
In the following we call this grid architecture $30^2$x$30$x$100 (3,3)$.
Such a grid takes about $300$~{\rm Mbyte}
of computer memory. The root-file where the grid is stored has about $50$~{\rm Mbyte}.

\begin{figure}[htbp]
\centering
\includegraphics[width=0.49\textwidth]{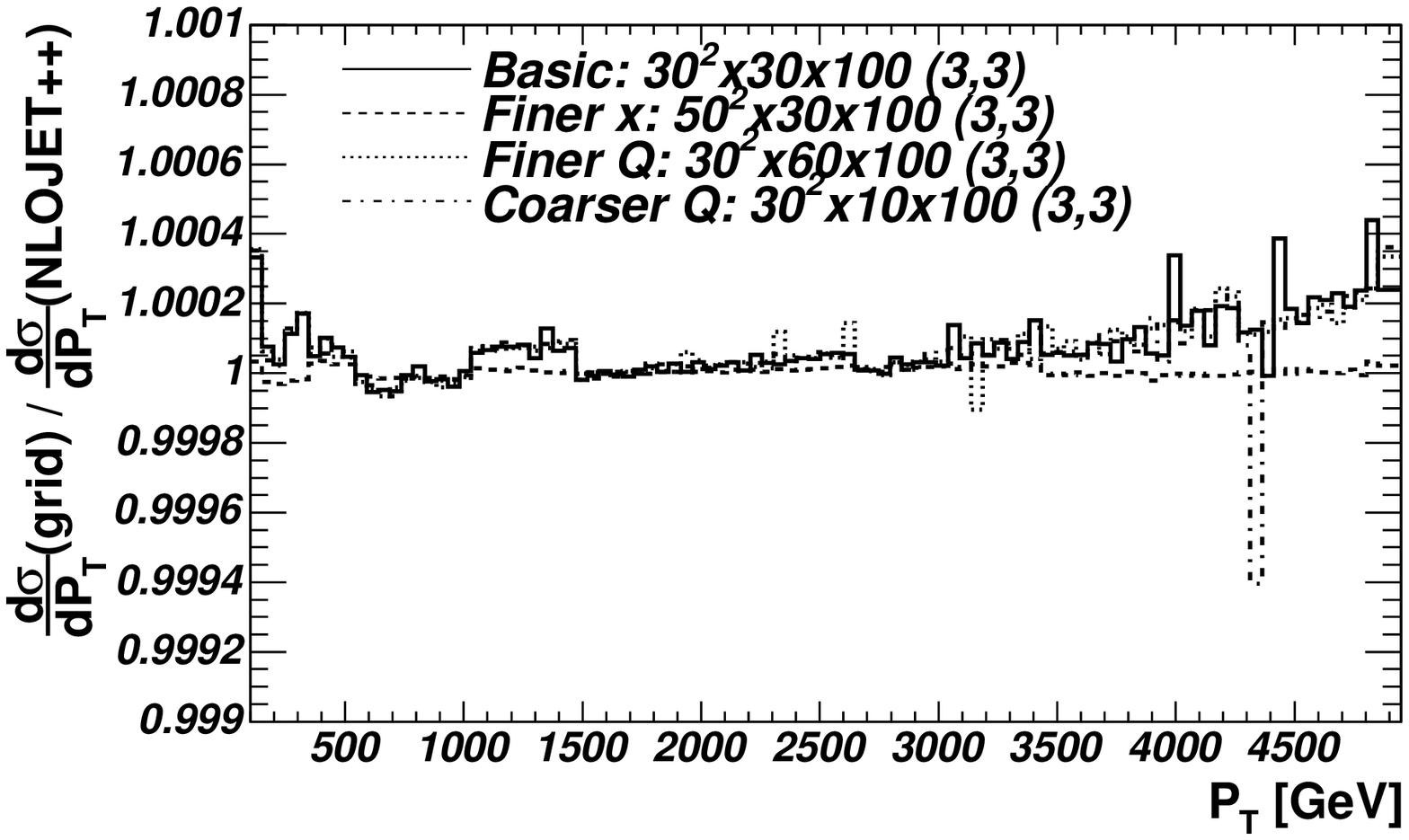}
\includegraphics[width=0.49\textwidth]{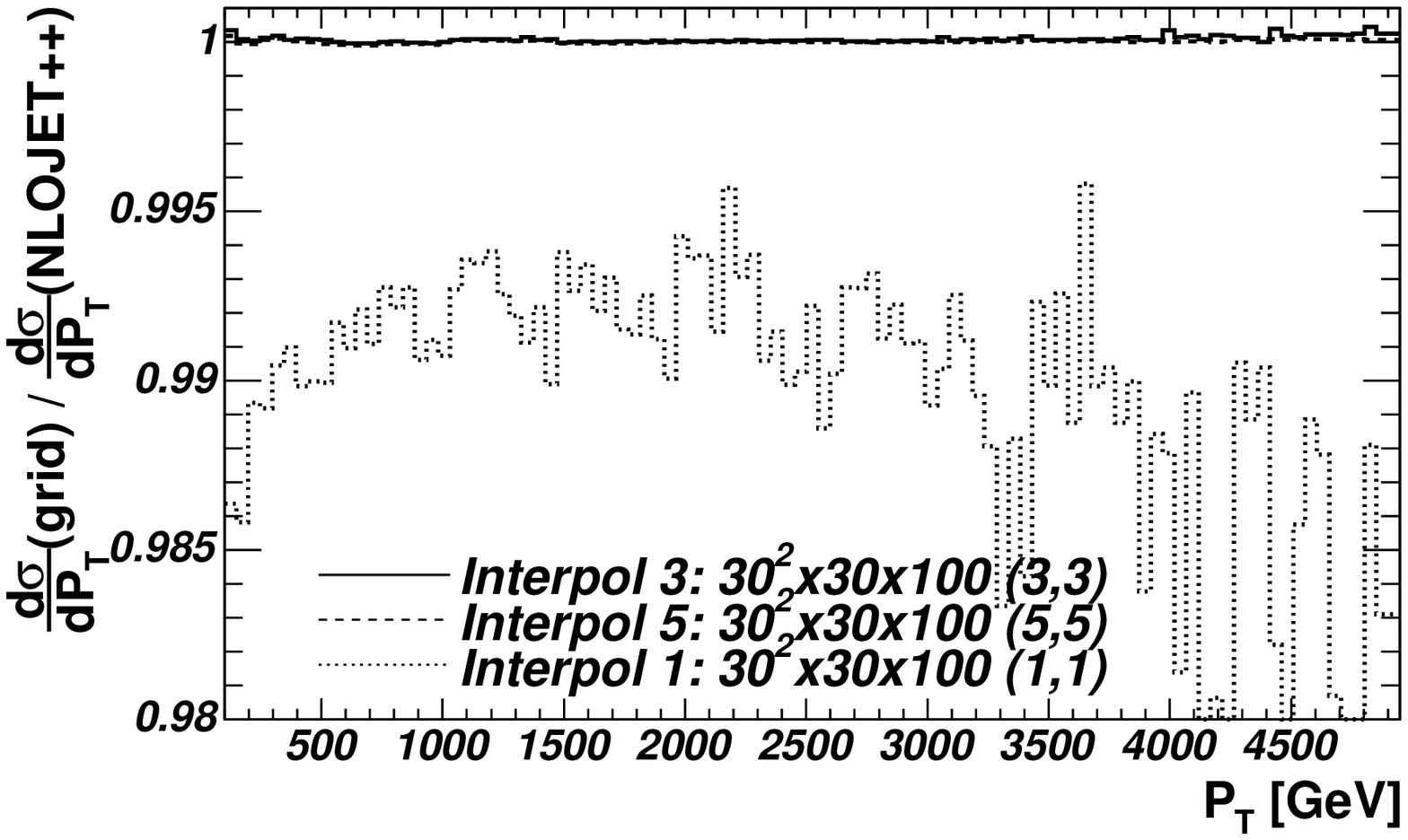}
\includegraphics[width=0.49\textwidth]{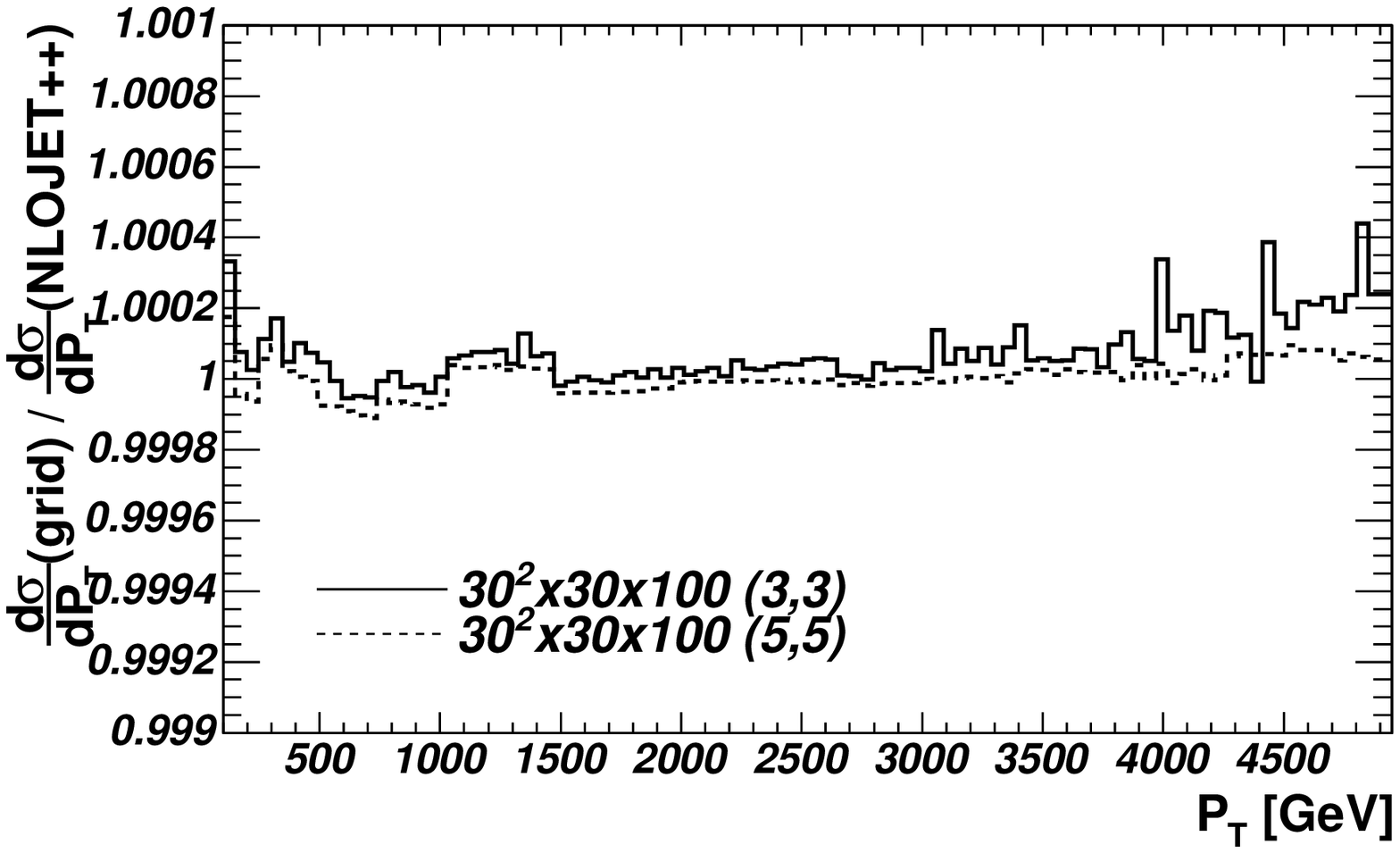}
\includegraphics[width=0.49\textwidth]{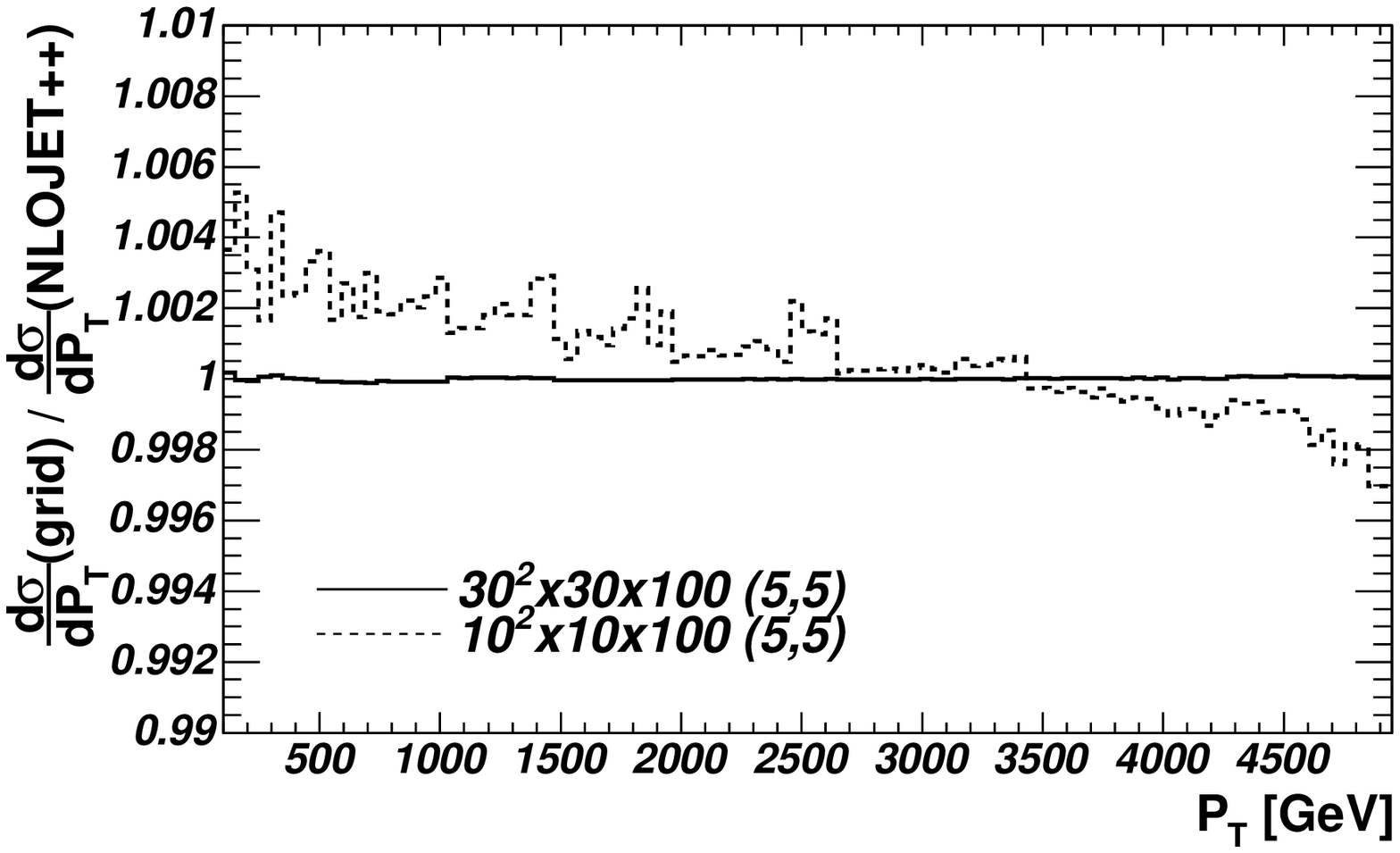}
\begin{picture}(0,0)
\put( -450,0){c)}
\put( -210,0){d)}
\put( -450,135){a)}
\put( -210,135){b)}
\end{picture}
\caption{Ratio between the single inclusive jet cross-section with $100$ $P_T$ bins calculated
with the grid technique and the reference cross-section calculated in the standard way.
Shown are the standard
grid, grids with finer $x$ and $Q^2$ sampling (a) with interpolation
of order $1$, $3$ and $5$ (b) (and on a finer scale in c)) and
a small grid (d).
}
\label{fig:101obsbins}
\end{figure}

The result is shown in Fig.~\ref{fig:101obsbins}a).
The reference cross-section is reproduced everywhere to within $0.05\%$. The typical precision is about $0.01\%$.
At low and high $P_T$ there is a positive bias of about $0.04\%$.
Also shown in Fig.~\ref{fig:101obsbins}a) are the results obtained with different grid architectures.
For a finer  $x$ grid ($50^2$x$30$x$100 (3,3)$) the accuracy is further improved (within $0.005\%$)
and there is no bias.
A finer ($30^2$x$60$x$100 (3,3)$) as well as a coarser  ($30^2$x$10$x$100 (3,3)$) binning in $Q^2$
does not improve the precision.

Fig.~\ref{fig:101obsbins}b) and Fig.~\ref{fig:101obsbins}c)
show for the grid ($30^2$x$30$x$100$) different interpolation methods.
With an interpolation of order $n=5$ the precision is $0.01\%$ and the bias at low and high $P_T$
observed for the $n=3$ interpolation disappears. The result is similar to the one
obtained with finer $x$-points. Thus by increasing the interpolation order
the grid can be kept smaller.
An order $n=1$ interpolation gives a systematic negative bias of about $1\%$
becoming even larger towards high $P_T$.

Depending on the available computer resources and the specific problem, the user
will have to choose a proper grid architecture.
In this context, it is interesting that a very small grid
$10^2$x$10$x$100 (5,5)$ that takes only about  $10$~{\rm Mbyte} computer memory
reaches still a precision of $0.5\%$, if an interpolation of order $n=5$ is used
(see Fig.~\ref{fig:101obsbins}d)).


\section{Conclusions}
We have developed a technique to store the perturbative coefficients calculated by an NLO
Monte Carlo program on a grid allowing for  a-posteriori inclusion of an arbitrary parton
density function (PDF) set. We extended a technique that was already successfully used to analyse
HERA data to the more demanding case of proton-proton collisions at LHC energies.

The technique can be used to constrain PDF uncertainties, e.g. at high momentum transfers,
from data that will be measured at LHC and allows the consistent
inclusion of final state observables
in global QCD analyses. This will help increase the sensitivity of LHC
to find new physics as deviations from the Standard Model predictions.

Even for the large kinematic range for the parton momentum fractions $x_1$ and $x_2$ and
of the squared momentum transfer $Q^2$ accessible at LHC, grids of moderate size
seem to be sufficient. The single inclusive jet cross-section in the central region $|y|<0.5$
can be calculated with a precision of $0.01\%$ in a realistic example with $100$ bins
in the transverse jet energy range $100\, \leq P_T \leq 5000\,\mathrm{GeV}$.
In this example, the grid  occupies about $300$~{\rm Mbyte} computer memory.
With smaller grids of order $10$~{\rm Mbyte}
the reachable accuracy is still $0.5\%$. This is probably sufficient for all practical
applications.

\section*{Acknowledgment}
We would like to thank Z. Nagy, M. H. Seymour, T. Sch\"orner-Sadenius, P. Uwer and  M. Wobisch
for useful discussions on the grid technique and A.~Vogt for
discussion on moment-space techniques. We thank Z. Nagy for help and support with
NLOJET++. F.~Siegert would like to thank CERN for the Summer Student Program.
%


\bibliographystyle{heralhc}
{\raggedright
\bibliography{carli}
}
\end{document}